\documentclass[prl,amsfonts,twocolumn]{revtex4}
\usepackage{amsmath,bm}    
\usepackage{amssymb}
\usepackage{amsthm}
\usepackage{bm}

\usepackage{amscd}
\usepackage{graphicx}   
\usepackage{verbatim}   
\usepackage{color}      
\usepackage{subfigure}  
\usepackage{hyperref}   
\usepackage{dsfont}

\usepackage{mathtools}

\newcommand{\Trace}{\operatorname{Tr}}

\DeclareMathOperator*{\Motimes}{\text{\raisebox{0.25ex}{\scalebox{0.8}{$\bigotimes$}}}}

\def \HilbL{\mathcal{H}}
\def \q{\mathbf{q}}
\def \ac{\mathbf{a}^{c}}
\def \Ue{\UI^{e}}

\def \Ucirc{U_{\mathrm{circ}}}
\def \Ugate{U_{\mathrm{gate}}}
\def  \Wo{\overline{W}}

\def \C{\mathcal{C}}

\def \TT{\mathcal{T}}
\def \T{\mathbf{T}}
\def \t{\mathbf{t}}

\def \WK{W_{\scriptscriptstyle  K}}
\def \WI{W_{\scriptscriptstyle  I}}

\def \HK{H_{\scriptscriptstyle  K}}
\def \HI{H_{\scriptscriptstyle  I}}

\def \UK{U_{\scriptscriptstyle  K}}
\def \UI{U_{\scriptscriptstyle I}}
\def \UN{U_{\scriptscriptstyle N}}

\def \TU{\tilde{U}}

\begin{document}
\title{Local  correlations  in dual-unitary  kicked chains }

\author{Boris Gutkin$^2$, Petr Braun$^1$, Maram Akila$^{1,3}$, Daniel Waltner$^1$, Thomas Guhr$^1$}
\affiliation{$^1$Fakult\"at f\"ur Physik, Universit\"at Duisburg-Essen,
  Lotharstra\ss e 1, 47048 Duisburg, Germany\\
  $^2$Department of Applied Mathematics, Holon Institute of Technology, 58102 Holon,
Israel\\
$^3$Fraunhofer IAIS, Schloss Birlinghoven, 53757 Sankt Augustin, Germany}

\begin{abstract}     We show  that for  dual-unitary  kicked chains,  built  upon a pair of complex Hadamard matrices,    correlators of strictly local, traceless  operators   vanish identically for  sufficiently long chains. On the other hand,  operators  supported  at  pairs   of adjacent chain sites, generically,   exhibit  nontrivial  correlations  along the light cone edges.  In agreement  with  Bertini et.~al.~[Phys.\ Rev.\ Lett.\ {\bf123}, 210601 (2019)], they can be expressed through the expectation values of a transfer matrix $\T$. Furthermore, we identify a remarkable  family  of dual-unitary models where an explicit information on the spectrum of $\T$ is available. For this class of models    we  provide a closed  analytical  formula  for the corresponding two-point   correlators. This result, in turn,  allows an evaluation of   local   correlators in the vicinity of the dual-unitary regime which is exemplified on the  kicked Ising spin chain.
\end{abstract}

\maketitle

 \section{Introduction}
 
 Spatially extended  Hamiltonian systems with local interactions provide  convenient frameworks for  theoretical  \cite{Engl2014, Dubertrand_2016, Abanin2015,Atas_2014,Keating2015, Czischek_2018} and experimental \cite{Schreiber842, Simon2011} studies in the field of many-body physics.
  Very generally, such systems  allow  a number of different dynamical descriptions \cite{AWGG16}. The standard    one  corresponds  to the system evolution  with respect to time,  induced by the system Hamiltonian.  Alternatively, one can consider  evolution along one of the   spatial directions. In such a case the corresponding coordinate takes on the role of  time. The resulting  dynamical system is generically a non-Hamiltonian one \cite{AWGG16, AWGBG16, AGBWG18}.  However,  in some special cases it might happen that the dual spatial evolution is a Hamiltonian one, as well. 
  The representatives  of such systems, referred to as \textit{dual-unitary}, can be found among coupled map lattices \cite{GutOsi15, GHJSC16}, kicked  spin chains \cite{BeKoPr18, BeKoPr19-1, BWAGG19, LakshPal2018}, circuit lattices \cite{BeKoPr19-4, GopLam19, BeKoPr2019operator}  and continuous field theories \cite{AVAN2016}. 
  
   Dual-unitary systems   have recently attracted   considerable attention \cite{BeKoPr19-4, GopLam19, BeKoPr19-1, LakshPal2018, BeKoPr18, BWAGG19,BeKoPrPi19, BeKoPr2019operator, KrPr2019KPZ,zhou2019entanglement,AVAN2016} due to their      intriguing properties.
  On the one hand,   these models generically exhibit features of  maximally chaotic many-body systems. In particular, their spectral statistics are   well described by the Wigner-Dyson distribution.  They  are  insusceptible to  many-body localisation  effects   even in the presence of strong disorder \cite{BeKoPr18, BWAGG19}. The  entanglement has been shown to grow linearly  with time and to  saturate   the maximum bound. 
  On the other hand,  dual-unitary  models turned out to be amenable to exact analytical treatment. The growth of the entanglement  entropy for kicked Ising spin chains (KIC) for certain types of initial states has been evaluated exactly in \cite{BeKoPr19-1} and their entanglement spectrum was found  to be trivial \cite{GopLam19}.   
  Furthermore, in  recent works of Prosen et. al. \cite{BeKoPr19-4}  it has been shown that two-point  correlations of strictly  local operators in  dual-unitary  quantum circuit   latices can be expressed exactly   in terms of small dimensional transfer matrices. 
  
  So far,  no full characterization of  dual-unitary  systems has been given. Although   concrete examples of such models have been presented, there is no general prescription for their construction. The present contribution aims to bridge  this gap. We introduce here  a wide class of dual-unitary kicked chains (DuKC) built upon a pair of $L\times L$ complex Hadamard matrices  and study  correlations between local operators. Importantly, these models are defined for arbitrary length of the chain $N$ and the on-site Hilbert  space dimension $L$. This allows, at least  in principle,  to look at both the thermodynamic, $N\to\infty$, and the semiclassical limit $L\to\infty$ (or combinations of them), which  is  important  for quantum chaos studies.     As  shown in the body of the paper,    the
  correlators  of strictly local traceless  operators vanish identically  in DuKC  for  sufficiently long chains. 
  On the other hand,   correlations  between   operators with finite support  are, generically,  non-trivial along the light-cone edges, in agreement  with the results of  \cite{BeKoPr19-4}. Such correlations can be expressed through the expectation values  of a  transfer matrix  $\T$  whose  dimension is determined by $L$ rather than $N$. 
  
  In what follows, we identify within DuKC a remarkable  family of  dual-unitary  models, where explicit information on the spectrum of $\T$ is available. For this family of DuKC we obtain  a closed analytical formula for correlations between operators supported on two adjacent lattice sites.   As a by-product,  this  allows an evaluation of correlations  between  local operators near  the dual-unitary regime, which is  illustrated on the example of KIC.  
  
\section{Dual-unitary  kicked chains} 
 In this paper we consider cyclic  chains of $N$ 
 locally interacting particles,  periodically kicked  with an on-site external potential.  The system is governed by the  Hamiltonian,
\begin{equation}
H= \HI+\HK\sum_{m=-\infty}^{+\infty}\delta(t-m),\label{KickedChain}
\end{equation}
with $\HI$, $\HK$  being the  interaction and kick parts, respectively.   
The corresponding  Floquet time evolution is the product  of the operators,  $\UI=e^{-i\HI}$ and   $\UK= e^{-i \HK}$,   acting on the Hilbert space  $\HilbL^{\otimes N}$  of  the dimension $L^N$, where    $\HilbL = \mathbb{  C}^L$ is  the local Hilbert space equipped   with the basis $\{|s\rangle, s=1,\dots, L\} $. We require  that   $\HI$ couples nearest-neighbour sites of the chain  taking on a   diagonal form   in the product basis,  $\{|\bm s\rangle=|s_1\rangle|s_2\rangle \dots|s_N\rangle\}$.  The respective evolution is   fixed by  a   real function $f_1$, 
\begin{equation}
\langle \bm s|\UI[f_1]|\bm{ s'}\rangle=\delta(\bm s ,\bm{s'}) e^{i \sum_{n=1}^N f_1(s_n,s_{n+1})},
\end{equation}
with $\delta(\bm s ,\bm{s'}) =\prod_{i=1}^N \delta(s_i-s_i')$, 
and  cyclic boundary condition $s_{N+1}\equiv s_1$.
The second, kick  part, is given by the tensor product
\begin{equation}
\UK[f_2]=\Motimes_{i=1}^N u_2,  \langle \bm s|\UK[f_2]|\bm{ s'}\rangle=\prod_{i=1}^N \langle s_i|u_2|{ s}'_i\rangle
\end{equation}
of the  local operator $u_2$.  Here  $u_2$ is a  $L\times L$  unitary  matrix whose   elements in the local basis  take the form
\begin{equation}
 \langle n|u_2|m\rangle=\frac{e^{i f_2(n,m)}}{\sqrt{L}}
\end{equation}
with $f_2$  being   in general a complex  function.
Combining the two parts together we obtain the quantum evolution 
\begin{equation}
U= \UI[f_1] \UK[f_2],
\label{eq:basePropagator}
\end{equation}
acting on the Hilbert space of dimension $L^N$.

In the same way one constructs the  dual evolution operator  acting  on the Hilbert space of dimension $L^T$ by exchanging $N\leftrightarrow T$ and $f_1 \leftrightarrow f_2$:  
 \begin{equation}
 \TU= \UI[f_2] \UK[f_1].
\end{equation}
 The following remarkable duality relation  \cite{AWGBG16,AGBWG18} holds between their traces for any integers $T$, $N$:
\begin{equation}
\mbox{Tr } U^{T}=\mbox{Tr }  \tilde{U}^{N}.
\end{equation} In contrast to the original evolution,  $\TU$ is a  non-unitary operator, in general. However, if   
\begin{equation}
\langle n|u_1|m\rangle=\frac{e^{i f_1(n,m)}}{\sqrt{L}},   \langle n|u_2|m\rangle=\frac{e^{i f_2(n,m)}}{\sqrt{L}},
\end{equation}
are  $L\times L$ unitary matrices which matrix elements have the same  absolute value, i.e.  $f_1(x,y) ,f_2(x,y)$  are real, the dual operator is unitary as well. We refer to such models as dual-unitary. 

It is a natural question to ask  how wide   the class of DuKC models is.  Each  dual model is  essentially built   upon a pair of  complex Hadamard matrices,   $u_1$ and $u_2$ (up to the $1/\sqrt{L}$ factor).  A generic family  of  complex Hadamard matrices can be constructed for each $L$ by taking the unitary discrete Fourier transform  (DFT) and multiplying it on both sides by diagonal unitary  and permutation matrices.    
This, however,  does not exhaust all possible cases. In general, the classification of complex Hadamard matrices is an  open problem \cite{Tadej2006}.

 
 \section{Strictly local    correlators}
 Let $\q_1, \q_2 $ be  a pair  of  traceless matrices acting on the on-site Hilbert space $ \HilbL$. We define the corresponding many-body operators
 \[Q_i=\underbrace{I\otimes  \dots\otimes I}_{n_i -1} \otimes\, \q_i\otimes \underbrace{I\otimes  \dots\otimes I}_{N-n_i }\]
  supported at the  $n_i$-th, $i=1,2$,  site of the chain, respectively.  Below we  show that  under the condition  $\max{(N- |n|,|n|)}>T$ with $ n= n_2 - n_1$ the correlator 
\begin{equation}
C_{1,2}=L^{-N}\Trace\left( U^T Q_1 U^{-T} Q_2\right) \label{two_point_correlator}
\end{equation}
vanishes for arbitrary  traceless $\q_1,\q_2$.
 This result  implies a lack of correlation between any pair of  operators $Q_1(z_1)=U^{-t_1} Q_1(n_1)U^{t_1}$,  $Q_2(z_2)=U^{-t_2} Q_2(n_2)U^{t_2}$,
 located at two different  points,
 $z_1=(n_1,t_1)$, $z_2=(n_2,t_2)$, of the spatial-temporal lattice. In other words, for sufficiently long chains, $N>|t_1-t_2|+|n_1-n_2| $, one has 
\begin{equation}
\langle  Q_1(z_1) Q_2(z_2)\rangle =\langle  Q_1\rangle \langle Q_2\rangle\label{two_point_correlator1}
\end{equation}
for $z_1\neq z_2$, where the average is defined as $\langle\cdot\rangle:= L^{-N}\Trace \,(\cdot)$.

 \textit{Dual representation.} To demonstrate that (\ref{two_point_correlator}) vanishes we will use the  dual  approach which allows us to rewrite correlators through the traces of operators acting in the dual space  $\HilbL^{\otimes 2T}$. Specifically, for the two-point correlator one has
\begin{equation}
C_{1,2}=\Trace \big(W^{N-n-1} \Wo_{{\q}_1} W^{n-1} W_{\q_2}\big),\label{two-point_correlator1}
\end{equation}
 if  $n= n_2-n_1\neq 0$ and 
\begin{equation}
C_{1,2}=\Trace \big(W^{N-1} W_{\q_1 \q_2} \big)\label{two-point_correlator2}
\end{equation}
 if  $n= 0$. The four dual evolution operators  are defined as follows:
\begin{equation}
\begin{split} &W_{\mathbf{b}} =\WI[\mathbf{1}, \mathbf{b}] \WK,   \, \overline{W}_{\mathbf{a}}=\WI[\ac, \mathbf{1}] \WK,\\ &
W=\WI[\mathbf{1}, \mathbf{1}] \WK,  \,  W_{\mathbf{a} \mathbf{b}} =\WI[\ac, \mathbf{b}] \WK,
\end{split}
\end{equation}
with $\ac = u_2\mathbf{a} u_2^\dagger$.  Similarly to the original time  evolution, the  dual operators are    products of kick and interaction parts.
The kick part $\WK$ has a tensor product structure
\begin{equation}\WK=\Motimes_{t=1}^{T} u_1^*\Motimes_{t=T+1}^{2T} u_1.\end{equation}
The interaction part  $W_I[\mathbf{a}, \mathbf{b}]$, defined for a pair of   local operators $\mathbf{a}, \mathbf{b}$,  takes on the form of  the diagonal matrix  in the basis  $\{|\bm s\rangle=|s_1\rangle|s_2\rangle \dots|s_{2T}\rangle\}$:
\begin{multline}
\langle \bm s |\WI[\mathbf{a}, \mathbf{b}]|{\bm s'}\rangle=
 \langle s_{2T}| \mathbf{a} |s_1\rangle\langle s_T| \mathbf{b} |s_{T+1} \rangle\delta(\bm s ,\bm{ s'})\\ e^{{-i \sum_{t=1}^{T-1} f_2(s_{t+1},s_{t}) + i \sum_{t=T+1}^{2T-1} f_2(s_t,s_{t+1})}}.
\end{multline}
In particular:
\begin{multline}
\langle \bm s |\WI[\mathbf{1}, \mathbf{1}]|{\bm s'}\rangle=
 \delta(  s_{2T}, s_1)\delta(s_T, s_{T+1}) \delta(\bm s ,{\bm s'})\\ e^{{-i \sum_{t=1}^{T-1} f_2(s_{t+1},s_{t}) + i \sum_{t=T+1}^{2T-1} f_2(s_t,s_{t+1})}}.
\end{multline}

\textit{Correlator evaluation.} Due to the presence of  $\WI$ the  operator $W$ is non-unitary. Further analysis shows that $W$ possess only one  non-zero eigenvalue $w_0=1$. In the dual case the right and left eigenvectors coincide  taking  the form:
\begin{equation}
|\Psi_0\rangle= {L^{-{T}/{2}}}\sum_{\bm s\in\HilbL^{\otimes T} } |{\bm s}\rangle \otimes \TT|{\bm s}\rangle,
\end{equation}
where $|{\bm s}\rangle =|s_1\rangle|s_2\rangle \dots|s_{T}\rangle$, and $\TT|{\bm s}\rangle= |s_T\rangle$ $|s_{T-1}\rangle \dots|s_{1}\rangle$.

\textbf{Proposition 1:} The matrix $W$ reduces to the rank-one projection after taking  the $T'$-th power, 
\begin{equation}
 W^{T'}= |\Psi_0\rangle\langle\Psi_0|, \label{Dualprojector}
\end{equation}
where $T'=T$ for even $T$ and  $T'=T+1$ for odd $T$, respectively.

\begin{proof}  We give the proof of (\ref{Dualprojector}) in the supplementary section of the paper. \end{proof}

\textbf{Remark:} An analogous  statement holds for non-dual case of kicked chain as well i.e., $f_2$ is a complex function. In general, $W^{T'}= |\Psi_R\rangle\langle\Psi_L|$, where $|\Psi_R\rangle$, $|\Psi_L\rangle$ are different $2T$-dimensional vectors. \\

Using  eq.~(\ref{Dualprojector})  the correlator can be reduced to the expectation value:
\begin{equation}
C_{1,2}=\langle \Psi_0|  \Wo_{{\q}_1} W^{n-1} W_{\q_2}| \Psi_0\rangle
\end{equation}
if $n \neq 0$ and 
\begin{equation}
C_{1,2}=\langle \Psi_0|  W_{\q_1\q_2}| \Psi_0\rangle 
\end{equation}
for $n=0$. 
The proof that $C_{1,2}$ in  (\ref{two_point_correlator}) vanishes  follows then immediately from the  proposition below.

\textbf{Proposition 2:}
For any traceless operator $\q$, holds:
\begin{equation}
W W_{\q} |\Psi_0\rangle=W \Wo_{\q} |\Psi_0\rangle=0.\label{BRSstate}
\end{equation}
\begin{proof} Let us first notice that   $ |\Psi_0\rangle $  stays invariant  under the action of $W_K$. This yields:
\[ W_{\q}|\Psi_0\rangle = \sum_{\bm s}\langle s_1|  \q| s_1\rangle \, |{\bm s}\rangle\otimes \TT| {\bm s}\rangle  \]
and analogously for $\Wo_{\q}|\Psi_0\rangle $.
It is then straightforward to check that an application to the last vector of $W$ leads to 
\[W W_{\q}|\Psi_0\rangle  = W \Wo_{\q}|\Psi_0\rangle  = \Trace \q |\Psi_0\rangle.\]
The last expression is obviously zero for traceless operators. \end{proof}

To obtain the factorization (\ref{two_point_correlator1}) it remains to notice that an arbitrary operator $Q_i$ can be split into the sum, \begin{equation}Q_i=\frac{\Trace Q_i}{L^N} \mathds{1}+Q'_i \label{decomposition}
\end{equation}
of  traceless $Q'_i$ and the unit operator $\mathds{1}$. Since the one and two point correlators,   $\langle Q'_1 Q'_2 \rangle$, $\langle Q'_1 \rangle$,  $\langle Q'_2 \rangle$ vanish, we arrive at (\ref{two_point_correlator1}). 
 This  result allows  a straightforward extension to $l$-point  correlators. 
 Let $Q(z_i)=\UN^{-t_i} Q_i   \UN^{t_i}$, $z_i=(n_i,t_i)$ be a set of  strictly  local   operators     supported at the ordered sites  $z_1, \dots,  z_l$  of the spacial-temporal  lattice i.e., $t_l > t_{l-1} >\dots >t_1$, $n_l > n_{l-1} >\dots >n_1$. 
As we show in the supplementary  material, for $N> |t_l-t_1| + |n_l-n_1| $ one has 
\begin{equation}
 \langle \prod_{i=1}^l Q_i(z_i) \rangle
 =\prod_{i=1}^l \langle Q_i \rangle, \label{ll_correlator}
\end{equation}
  under the condition that all operators are isolated from each other, i.e.   $|n_i-n_{i+1}|>1$ for all $i=1,\dots, l$.

\section{Operators with finite support  }
The condition 
that the operators $Q_i(z_i)$ are isolated is essential for (\ref{l_correlator}) to hold. As we show below,   operators supported on pairs of adjacent sites might have nontrivial  correlations along the light cone border.  Specifically,  
we consider here the time-ordered, $T>0$,  two-point correlator:
\begin{equation}
    C(n,T)=L^{-N}\Trace U^{T} \Sigma_0 U^{-T} \Sigma_n\label{eq:fourpointCorr}
\end{equation}
of the  operators 
\[\Sigma_0=Q_1(0)Q_2(1), \qquad  \Sigma_n=Q_3(n) Q_4(n+1)\]
localized at the points $0,1$ and $n,n+1$ respectively.  In the dual representation it takes on the form 
\begin{multline}
    C(n,T)=
    \Trace \overline W_{\q_1}  \overline{ W}_{\q_2} W^{n-1} {W}_{\q_3} W_{\q_4} W^{N-n-3}\\
  = \langle\Psi_0 |\overline{ W}_{\q_1} \overline{W}_{\q_2} W^{n-1} W_{\q_3} W_{\q_4}|\Psi_0 \rangle,
\end{multline}
where the last expression holds for sufficiently large $N$.
It is straightforward to check that $C(n,T)$ is zero if $n\neq T$. 
 As has been pointed out in \cite{BeKoPr19-4}, such   lack of correlations  can be understood in a simple intuitive way.    Due to the finite speed of information  propagation the two-point correlator  of two  traceless operators localized at the space-time lattice points $(0,0)$ and $(n,t)$, respectively,     must vanish  outside of the light cone $ t < |n|$.  By   the duality property, a similar result holds for  points within the light cone  $ t > |n|$, as well. This leaves the light cone edges $t=|n|$ as  the    only possible places on the space-time lattice  where non-trivial correlations might arise.  
 
As we show in the supplementary  material,  on the light cone edge  $C_T=C(T,T)$ does not vanish, rather it is given by the  expectation value
\begin{equation}
    C_T=\langle\bar{\Phi}_{\q_1\q_2}|\T^{T-2}|\Phi_{\q_3\q_4}\rangle,\label{four_point_corr0}
\end{equation}
of the transfer operator $\T$  acting on  the small  space $\HilbL\otimes \HilbL$.  The explicit form of the operator $\T$ and the corresponding vectors $\bar{\Phi}_{\q_1\q_2}, \Phi_{\q_3\q_4}$ are  provided in the supplementary section, see eqs.~(\ref{transferOp}, \ref{vectors1},  \ref{vectors2}). As  $\T$ is a doubly stochastic matrix, for typical system parameters  the correlators decay exponentially with the   rates determined  by the spectrum of $\T$.  In the next section we show that for a wide family  of   DuKC   the spectrum of $\T$, and the resulting correlators (\ref{four_point_corr0})   can be evaluated analytically.

\section{DFTC  model}
We recall that a DuKC model is fully determined by  the pair of the Hadamard matrices,  $u_1, u_2$.  The most straightforward way to  realize a kicked unitary-dual chain  is   to set $u_1=\Lambda_1 F \Lambda'_1$, $u_2=\Lambda_2 F \Lambda'_2$, where  $F$ is  $L\times L$  unitary DFT  and $\Lambda_1, \Lambda'_1, \Lambda_2, \Lambda'_2$ are arbitrary unitary diagonal  matrices with the elements $e^{i\lambda_1(m)}, e^{i\lambda'_1(m)}, e^{i\lambda_2(m)}, e^{i\lambda'_2(m)}$,  $m=1,2,\dots, L$. In such a case we have 

\[f_1(m,n)= -\frac{2\pi (m-1)(n-1)}{L}+ \lambda_1(m)+\lambda'_1(n),\]
\[f_2(m,n)= -\frac{2\pi (m-1)(n-1)}{L}+ \lambda_2(m)+\lambda'_2(n).\]
In what follows we  will refer to such models as  \textit{Discrete Fourier transform  chains} (DFTC). 

\textit{Eigenvalues.} By eq.~(\ref{transferOp}) (see supplementary material)   the elements of the transfer operator in the  DFTC  take  the form
\begin{equation*}
    \langle m n|\T |n'm' \rangle=\frac{1}{L^3}\left|\sum_{s=0}^{L-1} e^{\frac{2\pi i (m+n+m'+n'-4)s}{L}-i\mu(s+1)}\right|^2
\end{equation*}
where $\mu(s)=\lambda_1(s)+\lambda'_1(s)+\lambda_2(s)+\lambda'_2(s)$. Since the matrix elements depend only on the combination $m+n+m'+n'$,  $\T$ can be diagonalized  by using $F\otimes F$ unitary transformation. The resulting spectrum of $\T$  is composed  of $L$ non-trivial eigenvalues  supplemented by  $L(L-1)$  eigenvalues equal to $0$.  Explicitly, the non-trivial part of the $\T$ spectrum  is given by $\lfloor \frac{L-1}{2}\rfloor$ pairs of the   eigenvalues $\t_m=-\t_{L-m}= |d_m|,  m =1,2,\dots, \lfloor \frac{L-1}{2} \rfloor$, with 
\begin{equation}
    d_m=\frac{1}{L}\sum_{s=0}^{L-1} e^{i\mu(1+s)-i\mu(1+(s+m)\!\!\!\!\!\mod L) }, \label{DFTspectrum}
\end{equation} and  either one additional unpaired eigenvalue, $\t_0=1$, for odd  $L$,  or the two unpaired eigenvalues equal to $\t_0=1, \t_{L/2}=d_{L/2}$, for even $L$. 

\textit{Eigenvectors.}  
To construct the eigenvectors of $\T$ note that    $\Phi_{\mathbf{a}\mathbf{b}}$, $\bar{\Phi}_{\mathbf{a}\mathbf{b}}$  vectors are fixed by the choice of the local  operator  $ \mathbf{a}$, and the diagonal part of  $\mathbf{b}$, see eqs.~(\ref{vectors1}, \ref{vectors2}). Given  an integer $m$ let   $\mathbf{e}_m$  be the   diagonal matrix  with the    elements 
\[ \langle s|\mathbf{e}_m|s'\rangle=\delta(s,s')e^{-i2\pi s m/L}, \quad  s',s\in\{1,\dots, L\}.\]
It is straightforward to see that for an arbitrary $\mathbf{a}$  and $\mathbf{b}=\mathbf{e}_m$ the  corresponding vector   $\Phi_{\mathbf{a}\mathbf{e}_m}$ is an  eigenvector of $\T^2$ with the eigenvalue $|d_m|^2$. The eigenvectors of $\T$ are, therefore,  symmetric and antisymmetric combinations of    $\Phi_{\mathbf{a}\mathbf{e}_m}$ and $\Phi^*_{\mathbf{a}\mathbf{e}_m}$ for $ m =0,1,2,\dots, \lfloor L/2 \rfloor$:
\begin{equation}
\begin{aligned}
 |\Phi^s_{\mathbf{a},m}\rangle=e^{-i\phi_m/2} |\Phi_{\mathbf{a}\mathbf{e}_m}\rangle + e^{i\phi_m/2}|\Phi^*_{\mathbf{a}\mathbf{e}_m} \rangle,\\  |\Phi^a_{\mathbf{a},m}\rangle=e^{-i\phi_m/2}|\Phi_{\mathbf{a}\mathbf{e}_m}\rangle - e^{i\phi_m/2}|\Phi^*_{\mathbf{a}\mathbf{e}_m}\rangle, 
 \end{aligned}
 \label{DFTeigenvectors}
\end{equation}
$e^{i\phi_m}=\frac{d_m}{ |d_m|}$. They  correspond to the  eigenvalues $\t_m$ and  $\t_{L-m}$, respectively. Note that for $m=0$ and $m=L/2$ (for even $L$)  only symmetric  eigenvector exists. 

\textit{Correlators.} To obtain explicit form of the  correlator (\ref{four_point_corr0})
we decompose the  vectors $|\Phi_{\q_3\q_4}\rangle$ in the basis of the eigenstates. After application of $\T^{T-2}$ operators  this yields
\begin{equation}
C_T=\sum_{m=0}^{ \lfloor L/2 \rfloor} (\t_m)^{T-2}\left(2-\delta_{m,0}-\delta_{m,\frac{L}{2}}\right)\C_{m},
\end{equation}
where the coefficients $\C_{m}$ factorize in the products of four factors:
\begin{eqnarray}
\C_{m}&=&\mathrm{Re}[ e^{-i\phi}A^*_m(\q_4)A_m(\q^c_1) B^{(1)}_m(\q_3)B^{(2)}_m(\q^c_2)],\nonumber\\
\C_{m}&=&\mathrm{Re}[ A_m(\q_4)A_m(\q^c_1) B^{(1)}_m(\q_3)B^{(2)}_m(\q^c_2)]
\end{eqnarray}
for odd and even $T$, respectively. Here $A_m(\q)$ are defined  as DFT of the diagonal elements of $\q$:
\[ A_m(\q)=\frac{1}{L} \sum_{s=1}^{L} e^{i2\pi s m/L} \langle s|\q| s\rangle. \]
For the remaining factors one has 
\[ B^{(j)}_m(\q)=\frac{1}{L} \sum_{s=1}^{L}e^{i(\mu_j(s)-\mu_j(s^{(m)})) }\langle s|\q|s^{(m)} \rangle,\]
where $s^{(m)}=1+(s+m-1)\!\! \mod L$,  $\mu_1(s)=-\lambda_1(s)-\lambda'_1(s)-\lambda_2(s)$ and $\mu_2(s)=\lambda_1(s)+\lambda'_1(s)+\lambda'_2(s)$, respectively.   
For any  real observable $\q$ the relations  $A_m(\q)=A^*_{L-m}(\q)$,   $B^{(j)}_m(\q)=(B^{(j)}_m(\q))^*$,   $j=1,2$, hold for all $m$. Furthermore, for traceless $\q$ all factors vanish at $m=0$.

 \section{Application to    KIC} 
As we show in the  supplementary material, the self-dual KIC provides  a minimal,  $L=2$,  realisation of 
the DFTC  model  with the parameters
$
\mu(1)=-\pi/4  -h,  \mu(2)=-\pi/4+h.$

\textit{Strictly local  correlators.} By (\ref{two_point_correlator1})  it follows immediately that all possible two-point corelators  $\langle\sigma_n^\alpha(t)\sigma_m^\beta(0)\rangle$, $\alpha\neq\beta\in\{x,y,z\}$ between local spin operators vanish identically for $t>0$. As a simple corollary of this one obtains that the total magnetization $M^\alpha=\sum_n \sigma_n^\alpha$ has no correlations as well, i.e. $\langle M^\alpha(t) M^\beta(0)\rangle=0 $ for any combination of $\alpha, \beta$. 

\textit{Local  correlators.} In order to evaluate the correlation (\ref{eq:fourpointCorr}) between   operators with two site support we use eq. (\ref{four_point_corr0}). A straightforward calculation (see supplementary material) leads to
\begin{equation}
C( n, T)=\delta(n,T)\,\mathcal{C}_{\alpha\beta}^{\delta\gamma} \cos^n 2h ,\label{main_result}
 \end{equation}
 where the prefactors $\mathcal{C}_{\alpha\beta}^{\delta\gamma} $ depend  on the operators  $\Sigma_0=\sigma_0^\alpha\sigma_1^\beta,\Sigma_n=\sigma_n^\gamma \sigma_{n+1}^\delta$. Specifically,
 $
\mathcal{C}_{yz}^{yz}=1$, $ \mathcal{C}_{yx}^{xz} =\tan^2 2h$, 
$
 \mathcal{C}_{yx}^{yz}= 
 \mathcal{C}_{yz}^{xz} =-\tan 2h$
 and   zeroes for all other spin combinations. The  decay    of the correlators  (\ref{main_result}) is determined by the subleading  eigenvalue of $\T$ which is given by $\cos( \mu(1)-\mu(2))=\cos2h$ in accordance with eq.~(\ref{DFTspectrum}).  
 
\begin{figure}
    \includegraphics[width=0.4\textwidth]{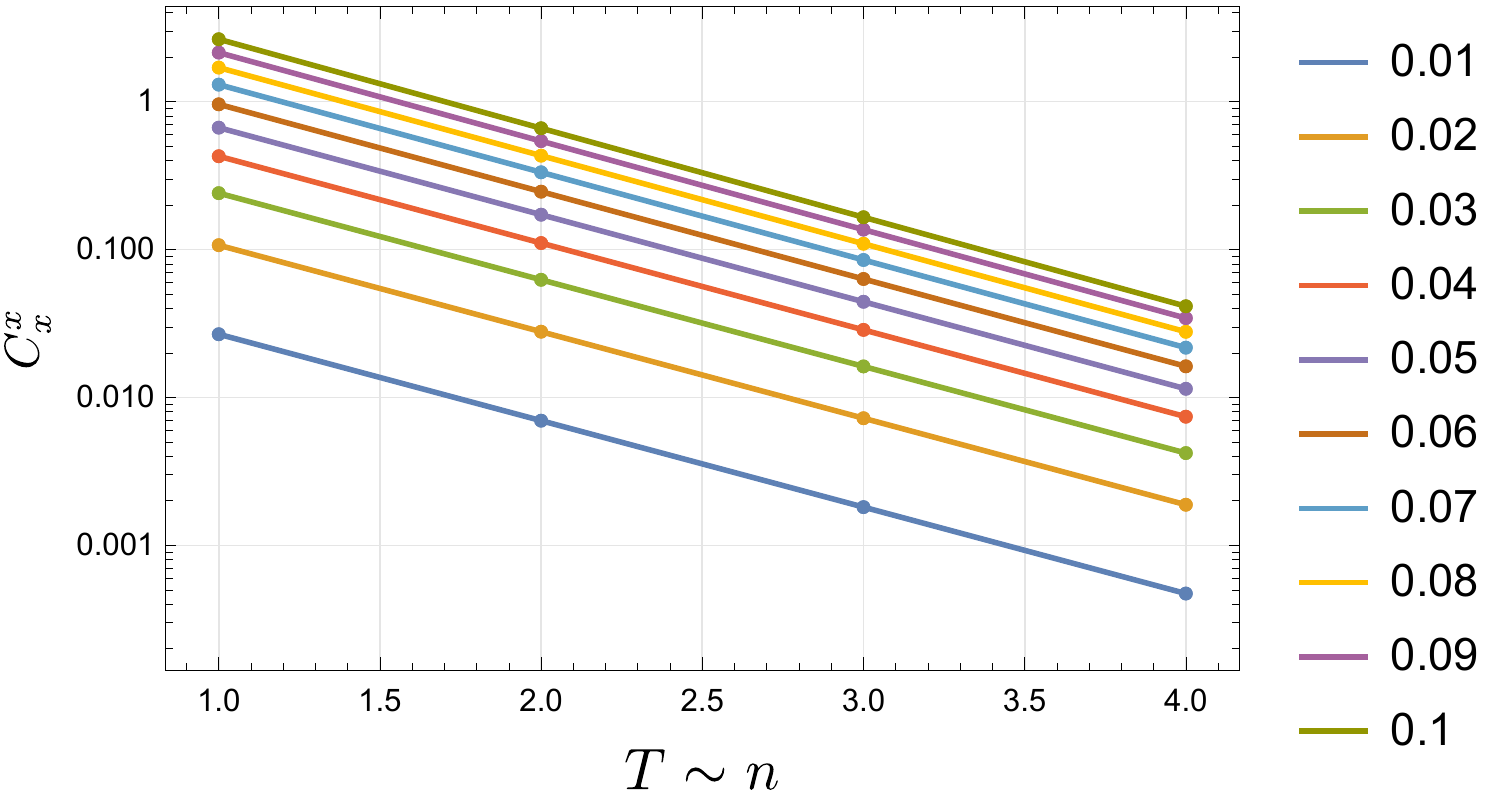}
    \caption[Decay of the correlator under perturbation of self-duality.]{The  behaviour of $C_{x}^x(n=T)=\Trace \big(U^{-T} \sigma_T^x {U}^{T}$ $ \sigma_1^x\big) $ for  $N=8$ spins with a generic value of $h=0.65362$. The system is near the self-dual line, i.e. $b=\pi/4$ and $J=\pi/4+\Delta J$. The  figure  shows the exponential decay with distance $n$, for various values of $\Delta J$, see legend. For $\Delta J \to 0$ the exponent is given by $\ln \cos{2h} \approx -1.35$, which holds in good approximation for all presented values of $\Delta J$.
    }
    \label{fig:corrDecayPert}
\end{figure}

\textit{Away from dual regime.} The above result can be used to evaluate the two point correlator away from the self-dual regime  in the leading order of perturbation. Indeed,  for $J=\pi/4 +\Delta J$ 
one has  to  the leading order of  $\Delta J$: 
 \begin{multline}C_{x}^x(n=T,  J)=\Trace \left(U_{ J}^{-T} \sigma_n^x {U}_{ J}^{T} \sigma_1^x\right)= \\
 =4(\Delta J)^2  \Trace \left(U^{-T} \Sigma_n {U}^{T} \Sigma_0\right) +O\left((\Delta J)^4\right),\end{multline}
 where $\Sigma_0 =\sigma_0^y \sigma_{1}^z$, $\Sigma_n =\sigma_n^y \sigma_{n+1}^z$ and $U$ is the  quantum evolution at ${\Delta J=0}$. By  the results on the  correlation function in the dual regime  we get in the leading order of perturbation an exponential decay,
 \begin{equation}
  C_{x}^x(n=T,  J)= 4(\Delta J)^2 \cos^T 2h  +O\left((\Delta J)^4\right)
 \end{equation}
 with the exponent given by $\ln \cos 2h$. The comparison with numerics is shown in fig.~\ref{fig:corrDecayPert}
 
\begin{figure}
    \includegraphics[width=0.4\textwidth]{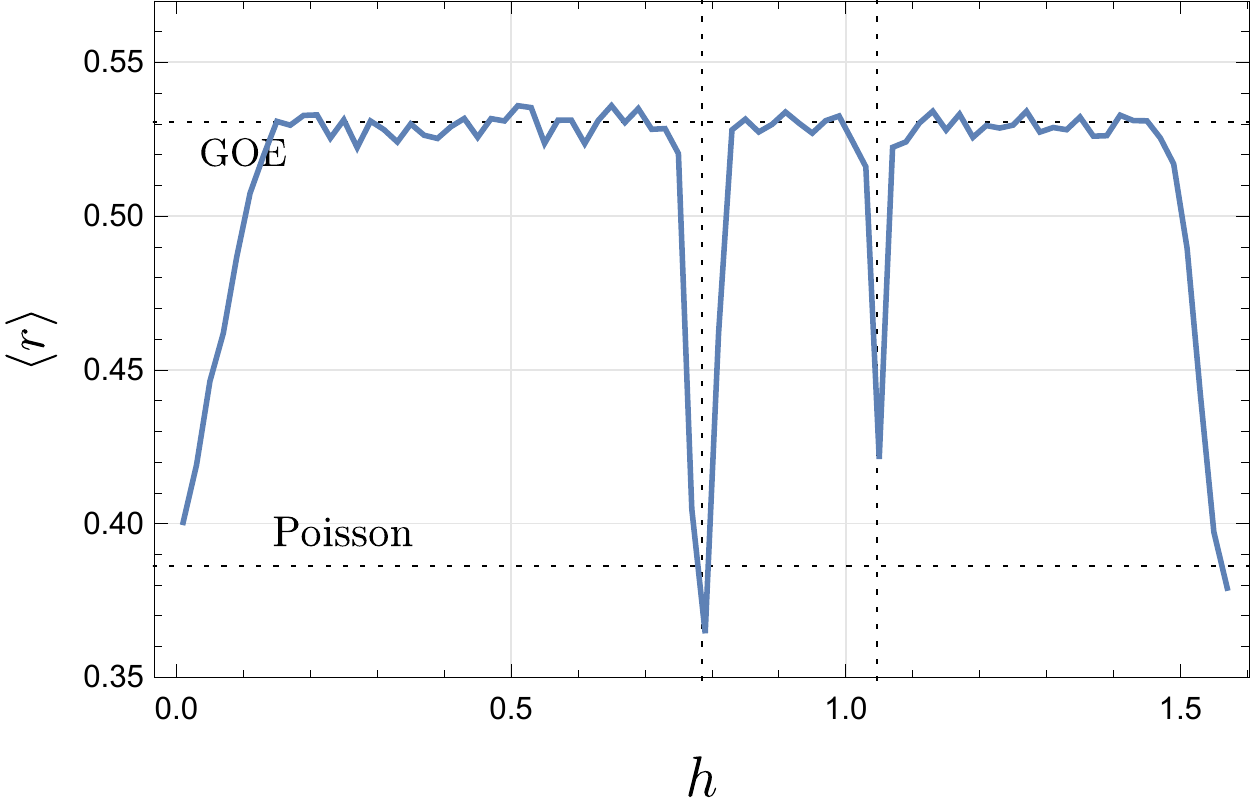}
    \caption[R-statistics.]{The figure shows the value of $r$ for the spectrum of the dual KIC at $N=15$ as a function of $h$ averaged over all values in the separate (translation) symmetry sectors of $k=1,\dots, 7$. Note that the graph is symmetric under reflection $h\to\pi-h$. For most values of $h$ the data for  $\langle r\rangle$ fit the GOE prediction. The four dips at $h=0,\pi/4, \pi/3, \pi$  correspond to  spectral statistics characteristic of integrable systems.}
    \label{fig:Rstatistics}
\end{figure}

\textit{Relation to spectral statistics.} By the translation symmetry spectrum of  KIC evolution operator can be split into $N$ uncorrelated subspectra $\{e^{i\theta^{(k)}_n}\}$,  $k=1,2, \dots, N$,  \cite{AWGG16}.  In fig.~\ref{fig:Rstatistics} we show 
  the averaged ratio  between three successive eigenphases   from the same sector, 
  \[r = \frac{\min \{ \theta^{(k)}_{n}-\theta^{(k)}_{n-1},\theta^{(k)}_{n+1}- \theta^{(k)}_{n}\} }
  { \max \{\theta^{(k)}_{n}-\theta^{(k)}_{n-1},\theta^{(k)}_{n+1}- \theta^{(k)}_{n} \}},\]
which is a well established  diagnostic for quantum chaos, see \cite{OganesyanHuse2007, Bogomolny2013}.

\begin{figure}
    \centering
    \includegraphics[width=0.48\textwidth]{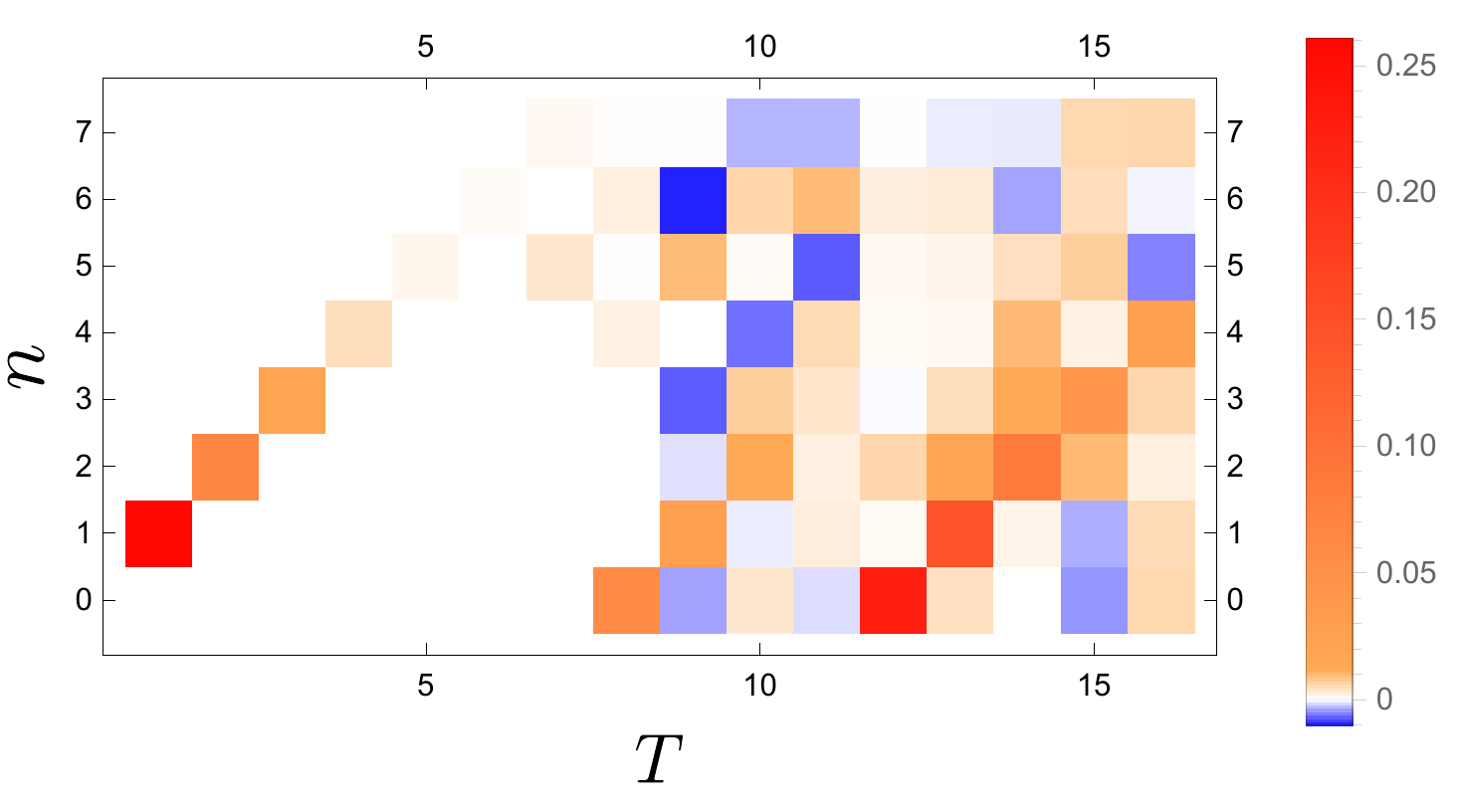}
    \caption[Non-local correlations on the self-dual line.]{Non-zero non-local correlations on the self-dual line given by $C_{zy}^{zy}(n,T)=\Trace {U}^{T} \sigma_1^z \sigma_0^y {U}^{-T} \sigma_{n+1}^z \sigma_{n}^y$. Visible is the exponential decay along one side of the light-cone, exchanging $y\leftrightarrow z$ exchanges the direction. After $T\geq N-1$ additional correlations, and a revival, occur. The shown system features $N=8$ and $h=0.65362$. }
    \label{fig:corrConeSD}
\end{figure}

For a generic value of $h$ the disymmetrized  spectrum of the self-dual KIC corresponds to a fully chaotic system. This is in agreement with the exponential decay of the   correlator (\ref{main_result}) on the light cone border, see fig.~\ref{fig:corrConeSD}.   There are, however, four  special points on the $h$-axis   where the KIC spectrum turns out to be ``non-chaotic".  The first three  points  $h=0, h=\pi , h=\pi/4$ correspond to  known cases of  the integrable classical 2-d Ising spin model \cite{LeeYangI,LeeYangII, Matveev_2008} with non-decaying  correlators (\ref{main_result}). 

The most intriguing is the last  ``integrable" case of $h=\pi/3$, which to the best of our knowledge has not been investigated   so far.  Here, despite Poissonian spectral statistics,  the correlators  decay exponentially, e.g.,   $C_{zy}^{zy}=  (-2)^{-T}$, on the light cone border (for $T<N-2$). This  is reminiscent of the  arithmetic surfaces of constant negative curvature, where correlations do decay exponentially, but the system spectrum exhibits Poissonian spectral statistics due to the existence of an infinite number of  Hecke operators  commuting with the system Hamiltonian \cite{Bogomolny1992}. In the same spirit we expect that for the self-dual KIC model at $h=\pi/3$ there exist an additional number of symmetries splitting the system's spectrum into uncorrelated  subspectra. Clarification of their exact nature is important, but beyond the scope of the present contribution. 


\section{Relation to dual-unitary circuit lattices}


For the sake of  comparison   it is instructive to observe a  connection  between the dual quantum kicked chain  and dual circuit lattices, studied in \cite{BeKoPr19-4}. Such a connection can be   established when both the  chain length $N$  and the propagation times $T$ are  even. It is straightforward to see that  the quantum evolution operator $U^{2T}$ for even times can be cast  into the form 
\begin{equation}
    U^{2T}=  \Ue \Ucirc^T (\Ue)^\dagger.
\end{equation}
Here, the operator $\Ue$ 
corresponds to the even  
half of the  "interaction":
\begin{equation}
\langle \bm s|\Ue[f_1]|\bm{ s'}\rangle=\delta(\bm s ,\bm{s'}) e^{i \sum_{n=1}^{N/2} f_1(s_{2n},s_{2n+1})},
\end{equation}
and the  evolution $\Ucirc$ has the form
\begin{equation}
\Ucirc= \mathbb{T} \Ue \UK \Ue \mathbb{T}^\dagger\Ue \UK\Ue,
\label{eq:circuitTrace}
\end{equation}
where $\mathbb{T}$ is the circular shift operator on a lattice of $N$ sites. Note that $\Ucirc$ has a special structure, 
characteristic to circuit lattice evolution \cite{BeKoPr19-4}. The role of the unitary gate operator is fulfilled here by 
\begin{equation}
\Ugate= u^e_1 \, (u_2\otimes u_2)\, u^e_1,
\end{equation}
where the diagonal matrix \[\langle  s_1 s_2|u^e_1|s'_1 s'_2 \rangle=\delta( s_1 ,{s'}_1)\delta( s_2 ,{s'}_2) e^{i  f_1(s_{1},s_{2})}\] 
is a restriction of $\Ue$ to two adjacent lattice sites.

By eq.~(\ref{eq:circuitTrace}) we find for two-point correlator
\begin{equation}
\Trace\left( U^T Q_1 U^{-T} Q_2\right)=\Trace\left( \Ucirc^T \tilde{Q}_1 \Ucirc^{-T}\tilde{Q}_1 \right), 
\end{equation}
where $\tilde{Q}_i= (\Ue)^\dagger Q_i\Ue$. Since $\Ue$ couples two neighbouring sites,  any local operator in the kicked model corresponds  to a two site operator of the respective  circuit model and vice versa.

 \section{Conclusions} We analyzed correlations between  local operators in DuKC   built  upon a pair of  $L\times L$ complex Hadamard matrices. The correlators of strictly local isolated traceless operators were shown to vanish identically for sufficiently long chains. On  the other hand  correlations between operators   with a finite support  were found to be, generically  non-trivial  along light-cone edges.    Here  an explicit formula, relating    correlators to the expectation values of a transfer operator has been derived. 
 For the subfamily of  DFTC  we  go much further and obtain  an explicit analytical expression  for correlations between operators  supported on  pairs  of adjacent sites.  Furthermore,  by using  these results   we  were able to evaluate     correlations between strictly local operators of KIC in the  vicinity of the dual regime. 

 So far, we have  discussed only  homogeneous   models. However, the results  of the current paper can be straightforwardly extended to dual-unitary  systems with spatial-temporal disorder. In such a case the transfer operator $\T^{n-2}$  is  substituted with  a product of  local ``gate" operators $\T_1\T_2\dots \T_{n-2}$, where each $\T_i$   depends  on  $f_1, f_2$ at the relevant point of the  spatial-temporal lattice. 
 For DFTC all matrices $\T_i$  are diagonalized by one and the same unitary transformation. As a result,  the decay exponents of the  correlators  (\ref{eq:fourpointCorr}) in the disordered case are just given by the averages of  the local exponents. In particular, for the non-homogeneous  KIC model   one has $C(n,T)\sim \delta(n,T) \prod_{i=1}^{n-2} \cos 2h_i$, where the $h_i$'s are  local magnetic fields at the corresponding points of the  spatial-temporal lattice. 
 
 
 Several open questions deserve further studies. First, a  possible   extension  of  the above results to all DuKC models should be explored. Second,  
 it would be of interest to investigate  whether    explicit results  for   correlations between  operators with larger supports  can be obtained for DFTC.  Finally,  the semiclassical limit $L\to\infty$ of DFTC  deserves a separate study. The   classical model emerging in this limit    is nothing more than  a  (perturbed) coupled cat map  lattice  considered in \cite{GutOsi15, GHJSC16}. Depending on  the functions  $\lambda_i(s), \lambda'_i(s)$  this model exhibits different  dynamical behaviours in the classical limit, ranging from full chaos to full integrability.  For a finite dimension $L$ of the local Hilbert space the two-point correlators of traceless operators  decay exponentially provided the transfer operator contains no eigenvalues on the unite circle, except the trivial one (associated with the unit operator). Thus, independently of the underlying classical dynamics,  for a fixed $L$  dual-unitary  systems  generically exhibit  quantum chaos behavior in the thermodynamic limit $N\to\infty$, associated with the exponential decay of correlators. On the other hand,  if the semiclassical limit $L\to \infty$ is taken first (or simultaneously with the thermodynamic limit)  the gap in the  transfer operator spectrum  might close, such that no exponential decay is observed for any finite $N$.   This shows that the emerging  theory is very sensitive to the order of the thermodynamic and semiclassical limits.

\vspace{1em} 
\section*{Acknowledgements} We thank  T.~Prosen for  useful discussion.  One of us (B.G.) acknowledges  support from  the Israel Science Foundation through grant No.~2089/19.

\bibliographystyle{ieeetr}
\bibliography{references_kicLim_Corr}


\phantom{p. 1}
\clearpage

\onecolumngrid
\section{Supplementary material}

\subsection{Dual representation of correlators}

As the first step, we  rewrite  correlator (\ref{two-point_correlator1}) in the form of  partition function for a classical statistical model. Specifically,  we have 
\begin{multline*}
    C_{1,2}=L^{-N}\Trace\left( U^T Q_1 U^{-T} Q_2\right)
    =\frac{1}{L^{NT}}
    \sum_{\{s_{m,t}\in 1,\dots, L\}}
    e^{-i\mathcal{F}(\{s_{m,t}\})}
    \langle s_{n_1,2T}| \q_1^c |s_{n_1,1}\rangle\langle s_{n_2,T}| \q_2 |s_{n_2,T+1}\rangle\\
    \times\prod_{m\neq n_1}^{N-1}\delta(s_{m,2T},s_{m,1})\prod_{m\neq n_2}^{N-1}\delta(s_{m,T},s_{m,T+1}),
\end{multline*}
where
\begin{multline}
\mathcal{F}=\sum_{t=1}^T \sum_{m=0}^{N-1} f_1(s_{m,t}, s_{m+1,t}) - f_1(s_{m,t+T}, s_{m+1,t+T})
    +f_2(s_{m+1,t}, s_{m,t}) - f_2(s_{m,t+T}, s_{m+1,t+T}).
\end{multline}
The last expression can be  rewritten through the transfer operators in the spatial direction as 
\begin{equation*}
C_{1,2}=\Trace \big(W^{N+n_1-n_2-1} \Wo_{{\q}_1} W^{n_2-n_1-1} W_{\q_2}\big).\label{two-point_correlator1_copy}
\end{equation*}
 if  $n_1\neq n_2$ and 
\begin{equation}
C_{1,2}=\Trace \big(W^{N-1} W_{\q_1 \q_2} \big)\label{two-point_correlator2_copy}
\end{equation}
 if  $n_1= n_2$.
 
 \begin{figure}
    \includegraphics[width=0.5\textwidth]{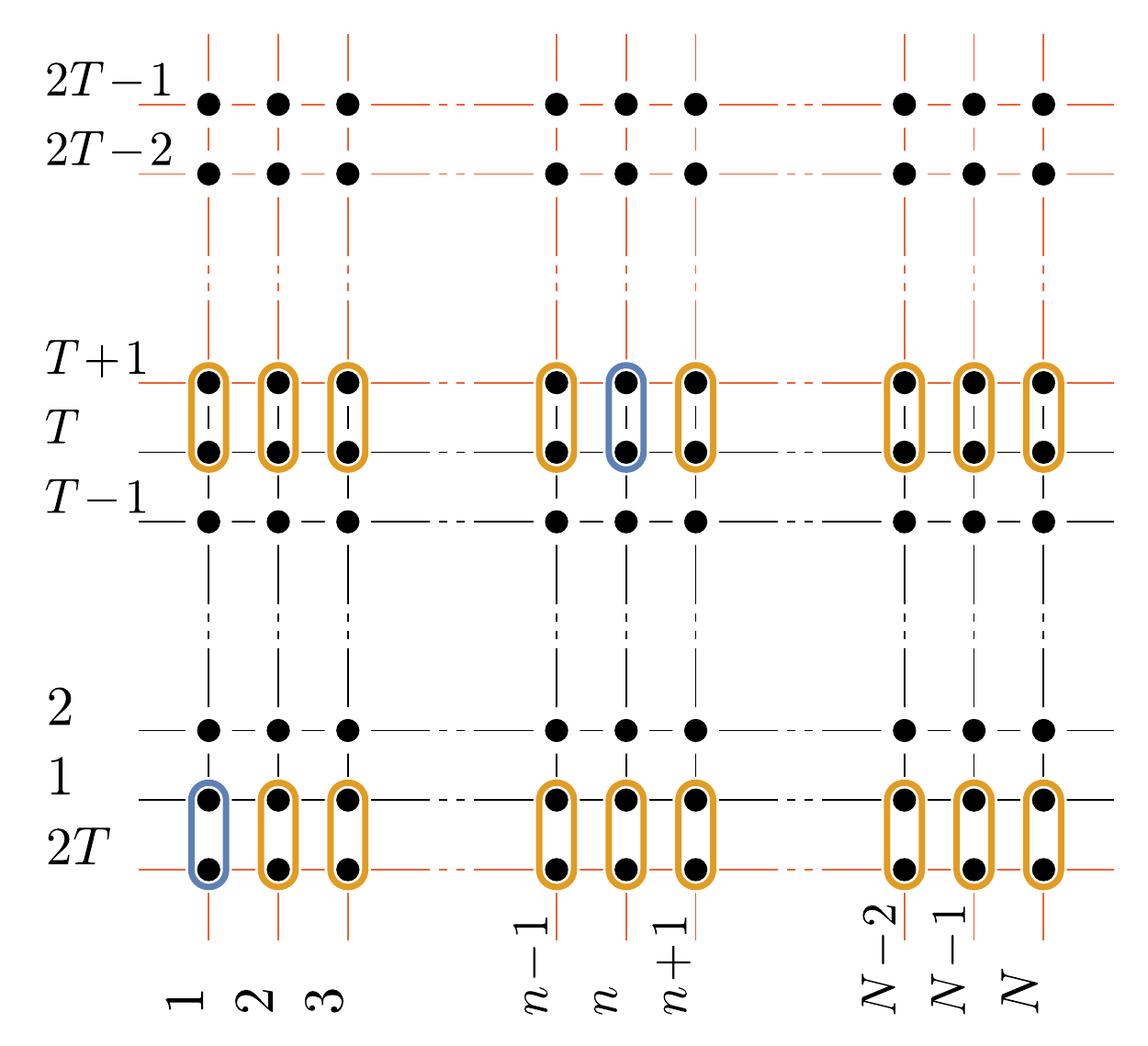}
    \caption[Decay of the correlator under perturbation of self-duality.]{The figure illustrates eq.~(\ref{two-point_correlator1}). The first and the n-th vertical line corresponds to the operators $\Wo_{\bar{\q}_1}$ and $W_{\q_2}$, respectively. The intermediate lines correspond to the operator $W$.  }
    \label{fig:DualCorrelator}
\end{figure}

 The  dual representation has a natural extension to  $l$-point correlator (\ref{l_point_correlator}). Assuming that all points are ordered, $t_l>t_{l-1}>\dots >t_1$,  $n_l>n_{l-1}>\dots >n_1$ we have
\begin{equation}
C_{1,\dots,l}=\Trace \big( W^{N+n_1
-n_l-1} W_1 W^{\Delta n_1}  W_2 W^{\Delta n_2} \dots  W_l\big),
\end{equation}
where $\Delta n_i=n_{i+1} -n_i -1$. Here
 $ W_1=  \Wo_{{\q}_1}, W_l= W_{{\q}_l}$, and  $W_k= \WI^{(k)}[\q_k]\WK$, for $1<k<l$ 
with
\begin{multline}
\langle \bm s |\WI^{k}[\mathbf{a}]|{\bm s'}\rangle=
 \langle s_k| \mathbf{a} |s_{k+1}\rangle
 \delta( s_1, s_{2T})\delta(s_T, s_{T+1}) \delta(\bm s ,{\bm s'}) e^{{\left(-i \sum_{t=1}^{T-1} f_2(s_{t+1},s_{t}) + i \sum_{t=T+1}^{2T-1} f_2(s_t,s_{t+1})\right)}}.\label{Def_Wk}
\end{multline}

\subsection{Proof of Proposition 1  }

In this section we are going to prove eq.~\ref{Dualprojector}. which in the matrix form  can be written as 
\begin{equation}
    \langle \bm \eta| W^T |\bm \eta'\rangle \sim \prod_{t=1}^T \delta(\eta_{t}- \eta_{T-t+1})\delta(\eta'_{t}- \eta'_{T-t+1}). \label{matrix_form}
\end{equation}
In order to prove this relation it is instructive to write down the left hand side of  (\ref{matrix_form}) in the form of partition function 
\begin{equation}
    \langle \bm \eta| W^T |\bm \eta'\rangle =
    \frac{1}{L^{NT}}\sum_{\{s_{n,t}\}}
    e^{i\left(\mathcal{F}_1(\{s_{n,t}\})+\mathcal{F}_2 (\bm\eta, \bm\eta')\right)} \prod_{n=1}^N\delta(s_{n,T}-s_{n,T+1})\delta(s_{n,2T}-s_{n,1}),\label{partition_form1}
\end{equation} 
 where the sum is over $NT$ variables $s_{n,t}$ with
\begin{equation}
    \mathcal{F}_1= \sum_{t=1}^T \sum_{n=1}^{N-1} f_1(s_{n,t}, s_{n+1,t}) - f_1(s_{n,t+T}, s_{n+1,t+T})
    +f_2(s_{n,t}, s_{n+1,t}) - f_2(s_{n,t+T}, s_{n+1,t+T})
    , \label{partition_form2}
\end{equation}
\begin{equation}
    \mathcal{F}_2= 
  \sum_{t=1}^T f_1(\eta_{t}, s_{1,t}) - f_1(\eta_{t+T}, s_{1,t+T})
 + f_1(\eta'_{t}, s_{N,t}) - f_1(\eta'_{t+T}, s_{N,t+T})
    . \label{partition_form3}
\end{equation}

\begin{figure}
    \centering

    \includegraphics[width=0.25\textwidth]{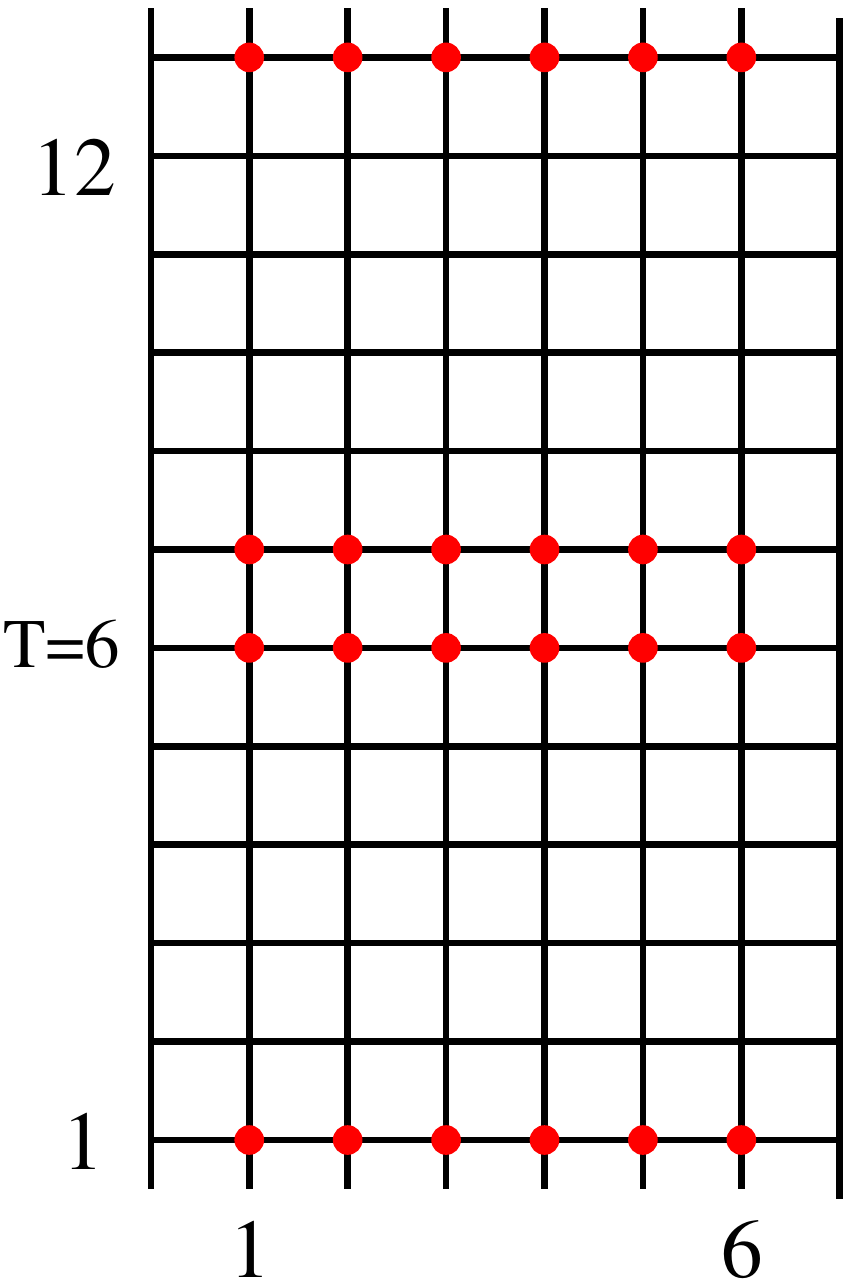}\hskip 2cm\includegraphics[width=0.25\textwidth]{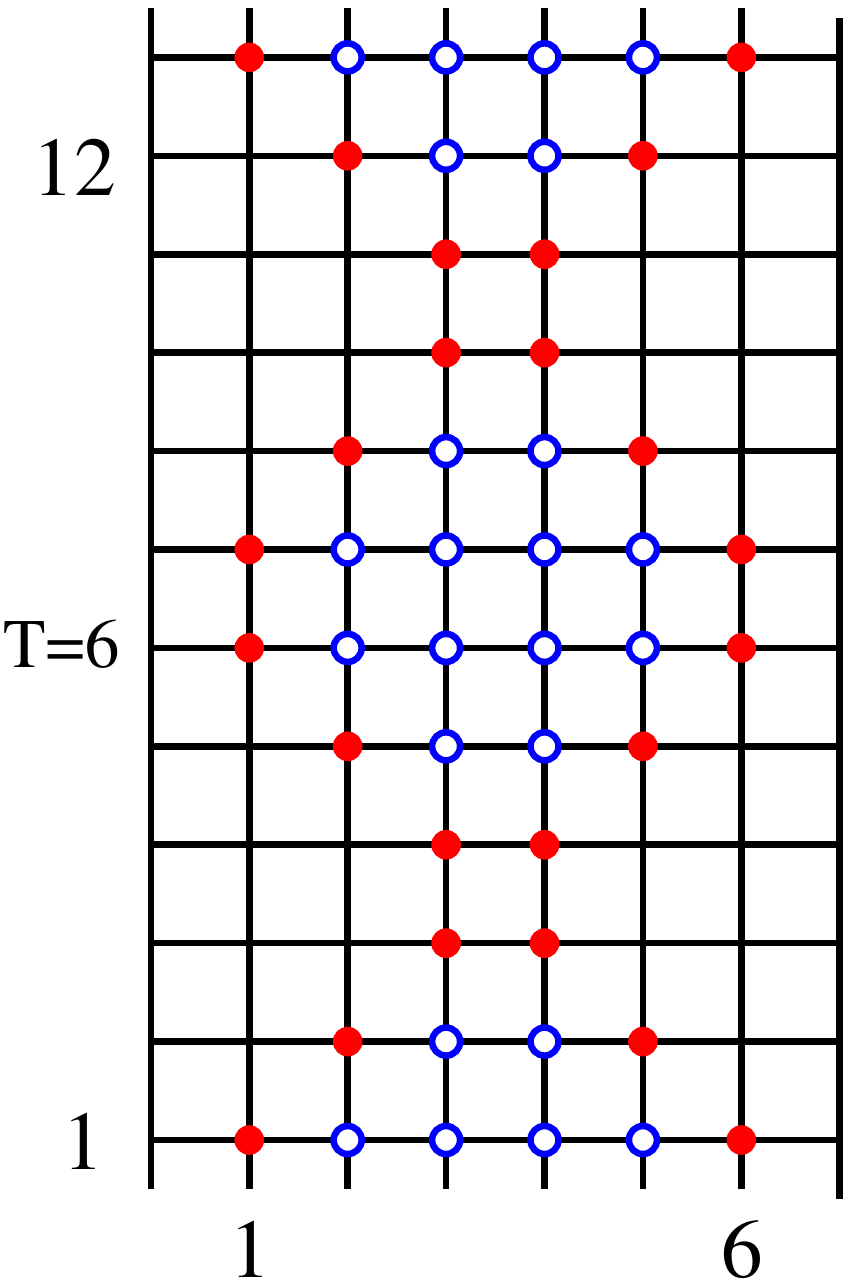}\hskip 2cm \includegraphics[width=0.25\textwidth]{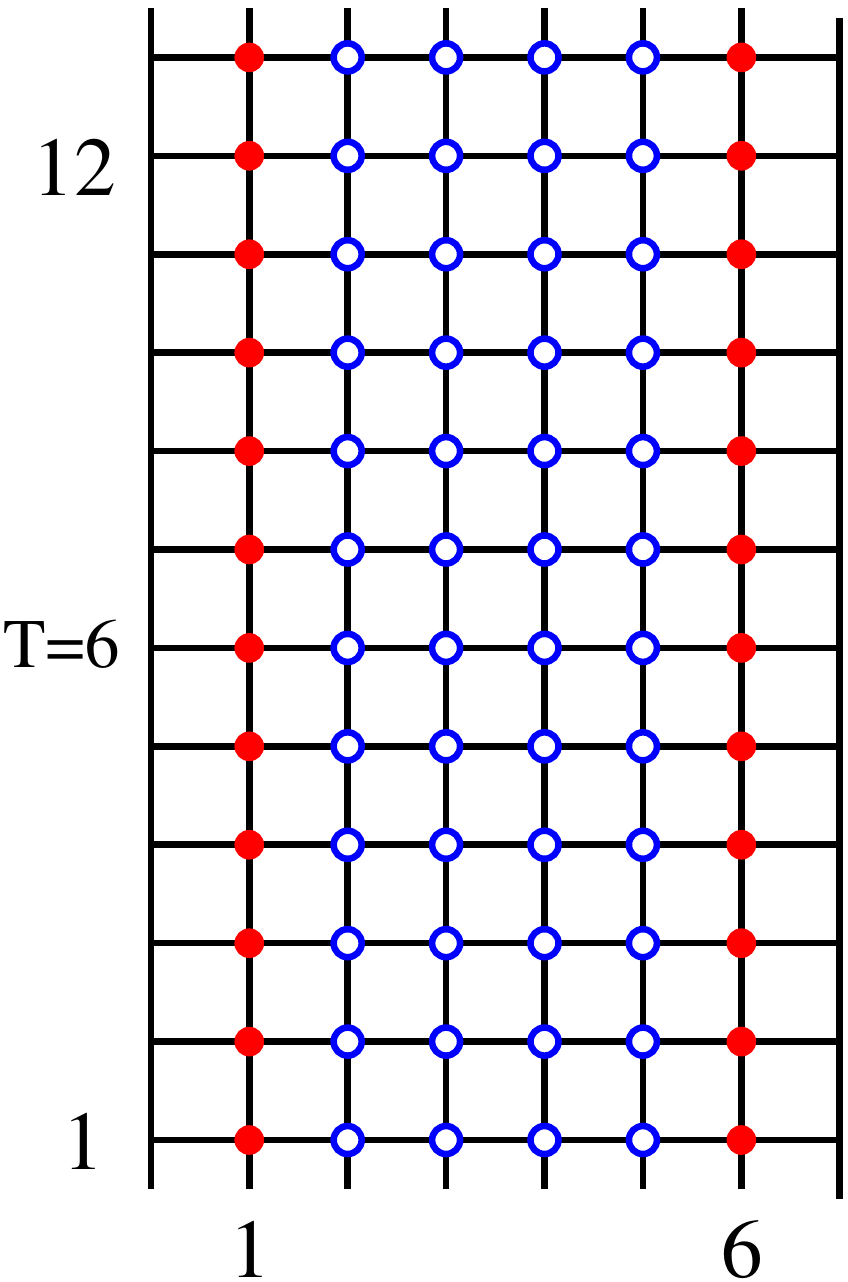}
    \caption[Non-local correlations on the self-dual line.]{Two step elimination process of the summation variables in the partition function. At the left picture is illustrated  the initial expression, where summation is ran over $N\cdot 2T$ variables $s_{n,t}$. Dots in red  show the spins which are paired by the condition  $s_{n,t}=s_{n,2T-t+1}$. At the first step all the  variables $s_{n,t}$  within the light cone are summed up (eliminated). They are  shown by the  blue (empty) circles in the middle figure. At the second step the variables $s_{n,t}$ outside of the light cone are eliminated, as illustrated at the right figure. }
    \label{fig:dual_grid}
\end{figure}

The summation  over the set of integers   $s_{n,t}$ is performed then in two steps. At first all the variables 
within  light cone are eliminated one by one as shown in fig. \ref{fig:dual_grid}. Then at the second step the summation variables $s_{n,t}$ outside of the light cone are eliminated except the first and the last raw. This  yields  
\begin{equation}
    \langle \bm \eta| W^T |\bm \eta'\rangle \sim
    \frac{1}{L^{NT}}\sum_{\{s_{1,t}, s_{N,t}\}}
    e^{i\mathcal{F}_2 (\{s_{1,t}, s_{N,t}, \eta_t,\eta_t' \})} \prod_{t=1}^N\delta(s_{1,t}-s_{1,2T-t+1})\delta(s_{N,t}-s_{N,2T-t+1}),\label{partition_form4}
 \end{equation} 
which after taking the sum  gives (\ref{matrix_form}).

 \subsection{\texorpdfstring{$l$}{l}-point    correlations between strictly local operators.}
 
 Let $Q(z_i)=\UN^{-t_i} Q_i   \UN^{t_i}$, $z_i=(n_i,t_i)$ be a set of   local traceless  operators     supported at the ordered sites  $z_1, \dots,  z_l$  of the spacial-temporal  lattice i.e., $t_l > t_{l-1} >\dots >t_1$, $n_l > n_{l-1} >\dots >n_1$. We are going to show that their correlations,
 \begin{equation}
C_{1,\dots,l}=L^{-N}\Trace \big( Q_1(z_1)Q_2(z_2) \dots Q_l(z_l)\big),\label{l_point_correlator}
\end{equation}
vanish identically for $N> |t_l-t_1| + |n_l-n_1| $, under the condition that all operators    $Q_i$ are    isolated from each other, i.e.   $|n_i-n_{i+1}|>1$ for all $i=1,\dots, l$.  
In the dual representation 
\begin{equation}
C_{1,\dots,l}=\Trace \big( W^{N-n_l
+n_1 -1} W_1 W^{\Delta n_1}  W_2 W^{\Delta n_2} \dots  W_l\big),
\end{equation}
where $\Delta n_i=n_{i+1} -n_i-1$. Here
 $ W_1=  \Wo_{{\q}_1}, W_l= W_{{\q}_l}$, and  $W_k$, $1<k<l$,  defined by  eq.~(\ref{Def_Wk}), have dimensions  $L^{2T}$, $T=t_l-t_1$. By applying  Propositions 1, 2 we get $C_{1,\dots,l}=0$ for traceless operators.

For general operators $Q_i$ we can again use the decomposition (\ref{decomposition}).  Under the condition that all operators are isolated and  spatial temporal ordered, one has 
\begin{equation}
 \langle \prod_{i=1}^l Q_i(z_i) \rangle
 =\prod_{i=1}^l \langle Q_i \rangle. \label{l_correlator}
\end{equation}

\subsection{Correlations between  operators with two-point support}

\begin{figure}
    \centering
    \includegraphics[width=0.25\textwidth]{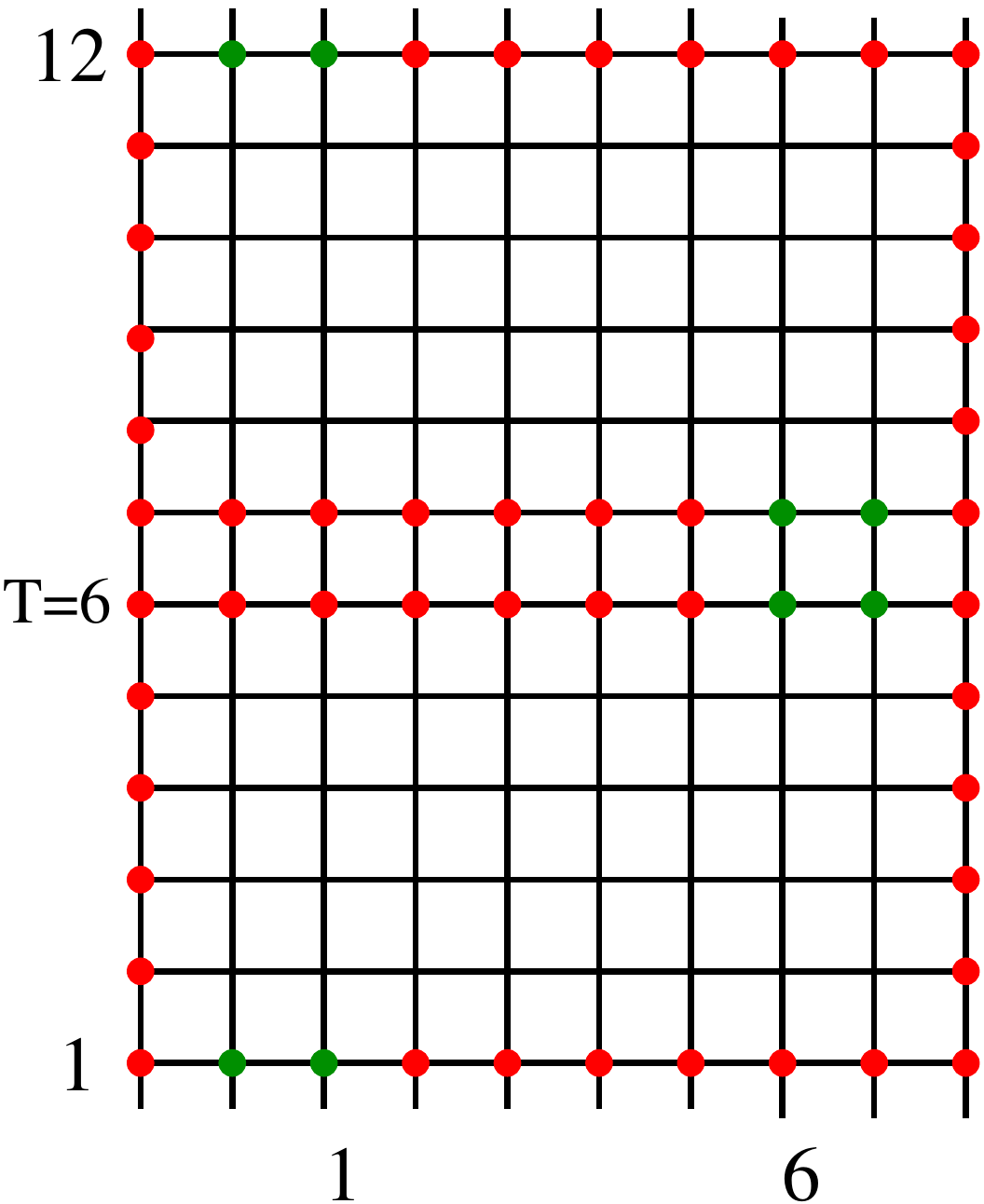}
    \hskip 3cm
    \includegraphics[width=0.25\textwidth]{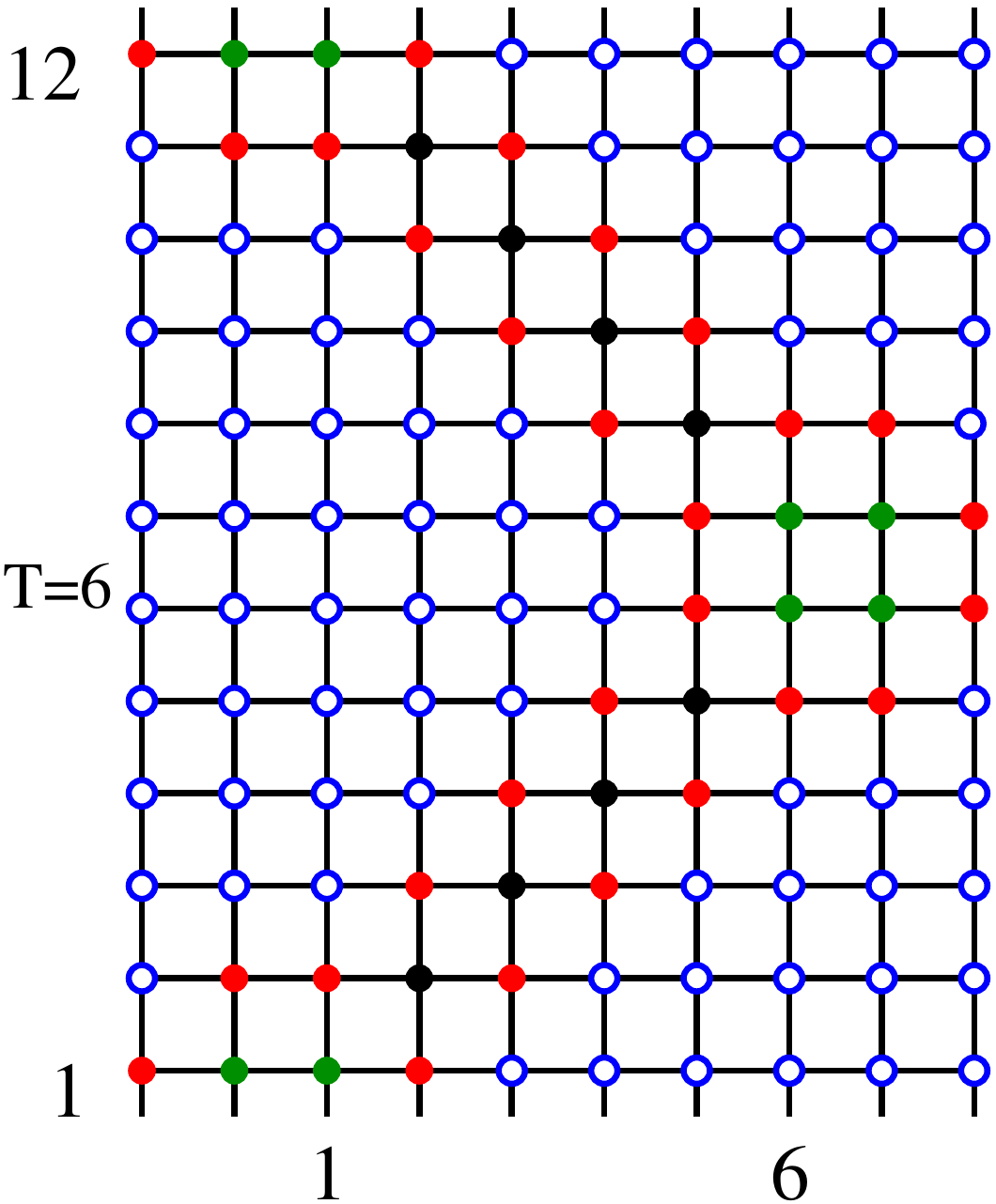}
    \caption[Non-local correlations on the self-dual line.]{Elimination  of the summation variables in the partition function representing 4-point correlator  for $T=6$. At the left picture is illustrated  the initial expression, where the sum  runs  over $(N+2)\cdot 2T$ variables $s_{n,t}$. Dots in red  show the variables  which are paired by the condition  $s_{n,t}=s_{n,2T-t+1}$. At the right picture is shown partition function after   elimination of  $s_{n,t}$, shown by blue circles. The remaining sum along the light-cone edge can be represented  in the form of the  expectation value (\ref{four_point_corr}) of the  $L^2\times L^2$  transfer  operator $\T$. }
    \label{fig:four_point}
\end{figure}
Here we consider the two point correlator of pairs of operators:
\begin{equation}
    C(n,T)=L^{-N}\Trace U^{T} \Sigma_0 U^{-T} \Sigma_n
\end{equation}
\[\Sigma_0=Q_1(0)Q_2(1), \qquad  \Sigma_n=Q_3(n) Q_4(n+1)\]
at the cone light border  $n=T$.
It is instructive to represent  $C_T=C(n=T,T)$  in the form of partition function. The initial expression   is shown in  a graphic form  on the left hand side of fig.~ \ref{fig:four_point}. The summation variables $s_{n,t}$ are excluded one by one up to reaching the stage illustrated by the right hand figure.  Here the summation variables (shown in red and black) are located along one dimensional strip only, which reduces the whole  problem to calculation of quasi-one dimensional partition function. Explicitly it can be cast into the form:
\begin{equation}
    C_T=\langle\bar{\Phi}_{\q_1 \q_2}|\T^{T-2}|\Phi_{\q_3 \q_4}\rangle,\label{four_point_corr}
\end{equation}
where the left $\bar{\Phi}_{\q_1\q_2}$ and the right  $\Phi_{\q_3\q_4}$  vectors are defined as 
\begin{eqnarray}
\langle\nu \eta|\Phi_{\q_3\q_4}\rangle&=&
\frac{1}{L^3}\sum_{a,\bar{a},b=1}^Le^{i(f_1(\eta,\bar{a})-f_1(\eta, a) +f_2(\bar{a},\nu)-f_2(a,\nu)-f_1(a,b)+f_1(\bar{a},b))}
 \langle a|\q_3|\bar{a} \rangle \langle b|\q_4|b\rangle, \label{vectors1} \\
\langle\bar{\Phi}_{{\q}_1{\q}_2}|\eta\nu\rangle&=&
\frac{1}{L^3}\sum_{a,\bar{a},b=1}^Le^{i(f_1(a,\nu)-f_1(\bar{a},\nu) +f_2(\eta,a)-f_2(\eta,\bar{a})+f_1(b,a)-f_1(b,\bar{a}))}
 \langle a|u_2 {\q}_2 u^\dagger_2|\bar{a} \rangle \langle b|u_2 {\q}_1 u^\dagger_2|b\rangle, \label{vectors2}
\end{eqnarray}
and
the transfer operator $\T$,
\begin{equation}
   \langle\nu \eta|\T|\eta'\nu'\rangle=
   \frac{1}{L^3}\left|\sum_{s=1}^L e^{i(f_1(\eta,s)+f_1(s,\nu') +f_2(\nu,s)+f_2(s,\eta'))}\right|^2 ,\label{transferOp}
\end{equation}
  acting on  the small  space $\HilbL\otimes \HilbL$.
  
It is easy to check that $\T$ is doubly stochastic operator i.e, satisfies
\begin{equation}
 \sum_{\nu=1}^L\sum_{\eta=1}^L\langle\nu \eta|\T|\eta'\nu'\rangle=1.
\end{equation}
This implies  that the  spectrum of $\T$ is contained within the unit disc with   at least one eigenvalue equal to $1$ corresponding to uniform eigenvector.

\subsection{Application to  DFTC model}

By eq.~(\ref{four_point_corr}) we have 
\begin{equation}
C_T=\sum_{m=0}^{ \lfloor L/2 \rfloor} (\t_m)^{T-2}\left(2-\delta_{m,0}-\delta_{m,\frac{L}{2}}\right)\mathrm{Re}[ A_m(\q_4) \langle\bar{\Phi}_{\q_1\q_2}|\Phi_{\q_3\mathbf{e}_m}\rangle], \label{Morecorrelators1}
\end{equation}
 for even $T$ and
\begin{equation}
C_T=\sum_{m=0}^{ \lfloor L/2 \rfloor} (\t_m)^{T-2}\left(2-\delta_{m,0}-\delta_{m,\frac{L}{2}}\right)\mathrm{Re}[ e^{-i\phi}A^*_m(\q_4) \langle\bar{\Phi}_{\q_1\q_2}|\Phi_{\q_3\mathbf{e}_m}\rangle],\label{Morecorrelators2}
\end{equation}
  for odd $T$, where 
  \[ A_m(\q)=\frac{1}{L} \sum_{s=1}^{L} e^{i2\pi s m/L} \langle s|\q| s\rangle.\]
The scalar products $\langle\bar{\Phi}_{\q_1\q_2}|\Phi_{\q_3\mathbf{e}_m}\rangle$ can be easily evaluated by using eqs.~(\ref{vectors1}, \ref{vectors2})
\begin{equation}
\langle\bar{\Phi}_{\q_1\q_2}|\Phi_{\q_3\mathbf{e}_m}\rangle = A_m(\q^c_1) B^{(1)}_m(\q_3)B^{(2)}_m(\q^c_2),\quad  B^{(j)}_m(\q)=\frac{1}{L} \sum_{s=1}^{L}e^{i(\mu_j(s)-\mu_j(s^{(m)})) }\langle s|\q|s^{(m)} \rangle,\quad  j=1,2
\end{equation}
with $s^{(m)}=1+(s+m-1)\!\! \mod L$, and $\mu_1(s)=-\lambda_1(s)-\lambda'_1(s)-\lambda_2(s)$ and $\mu_2(s)=\lambda_1(s)+\lambda'_1(s)+\lambda'_2(s)$, respectively. Note that the constants $ A_m(\q), B^{(j)}_m(\q)$  can be also written in  a more compact form as
\begin{equation}
A_m(\q)=\frac{1}{L}\Trace\left(\Gamma^m_0 \q\right),\qquad  B^{(j)}_m(\q)=\frac{1}{L}\Trace\left(\Gamma_{j} \q \Gamma_{j}^\dagger \,\mathbb{T}^m\right),
\end{equation}
where $\mathbb{T}$ is the circular shift operator,   $\mathbb{T} |s\rangle= |s^{(m)}\rangle$, and $\Gamma_j, j=1,2,3$ are the  diagonal matrices:
\begin{equation}
\Gamma_0=\mbox{diag}\{e^{i2\pi s/L}\}_{s=1}^{L}, \qquad \Gamma_j=\mbox{diag}\{e^{i\mu_j(s)} \}_{s=1}^{L}, \qquad  j=1,2.   
\end{equation}

\subsection{Application to KIC model}

The KIC model provides  a minimal realisation of self-dual models with $L=2$. The  KIC evolution is governed by the  Hamiltonians:
\begin{equation}
 \HI=\sum_{n=1}^N J\sigma_n^z \sigma_{n+1}^z +h\sigma_n^z,\qquad 
\HK=b\sum_{n=1}^N  \sigma_n^x,   \label{KICHamiltonian} 
\end{equation}
where $\sigma_n^\alpha, \alpha=x,y,z$ are Pauli matrices.  The dual case  corresponds to $J=b=\pi/4$ with $h$  being arbitrary.  
The  resulting  evolutions  $\UK$, $\UI$  take the form  (\ref{eq:basePropagator}) with the functions  
\[f_1=-\frac{\pi}{4} mn - \frac{h}{2}(m+n),  \qquad
f_2=\frac{\pi}{4} (mn-1) , \]
$m,n=\pm 1$, defining the two unitary matrices $u_1,u_2$:
\begin{equation}
    u_1=\frac{1}{\sqrt{2}}\begin{pmatrix} 
  e^{-i(\frac{\pi}{4}+h)} &   e^{i\frac{\pi}{4}}   \\
  e^{i\frac{\pi}{4}} &   e^{-i(\frac{\pi}{4}-h)} 
\end{pmatrix}, \qquad  u_2=\frac{1}{\sqrt{2}}\begin{pmatrix} 
  1 &  -i  \\
-i &  1 
\end{pmatrix}. \label{u1u2KIC}
\end{equation}
Note that $u_1,u_2$ can be expressed through the DFT matrix $F$ as:
\begin{equation*}
u_1=\begin{pmatrix} 
  e^{-\frac{ih}{2}} &   0   \\
 0 &   e^{\frac{i(\pi+h)}{2}} 
\end{pmatrix} F 
\begin{pmatrix} 
  e^{-\frac{i(\pi+2h)}{4}} &   0   \\
 0 &   e^{\frac{i(\pi+2h)}{4}} 
\end{pmatrix}
, \quad u_2=\begin{pmatrix} 
 1 &   0   \\
 0 &   e^{-\frac{i\pi}{2}} 
\end{pmatrix} F 
\begin{pmatrix} 
  1 &   0   \\
 0 &   e^{-\frac{i\pi}{2}} 
\end{pmatrix},
\quad   F=\frac{1}{\sqrt{2}}\begin{pmatrix} 
  1 &  1  \\
1 &  -1 
\end{pmatrix}. \label{u1u2Fourier}
\end{equation*}
This implies that KIC  is just a particular case of the  DFTC model for $L=2$ with the parameters  
 \begin{equation}
\Lambda_1=\mbox{diag}\{e^{-i{h}/{2}},e^{i(\pi+h)/2} \},\quad \Lambda'_1=\mbox{diag}\{e^{-i(\pi +2h)/{4}},e^{i(\pi+2h)/4}\},\quad \Lambda_2=\Lambda'_2=\mbox{diag}\{1,e^{-i\pi/2}\}. 
\end{equation}

For the  KIC model both the transfer operator  (\ref{transferOp}) and  the vectors (\ref{vectors1}, \ref{vectors2}) can be   calculated explicitly.  Inserting into eq.~(\ref{transferOp}) the corresponding functions $f_1, f_2$
yields:

\begin{equation}
    \T=\frac{1}{2}\begin{pmatrix} 
  a &  b  & b  & a\\
b &  a  & a & b \\
b &  a  & a & b \\
a &  b  & b  & a
\end{pmatrix}, \label{transferKIC}
\end{equation}
where $a=\cos^2 h$,   $ b=\sin^2 h$. The four eigenvalues of this matrix are $\{1,\cos 2h, 0, 0\}$ in agreement with the results of \cite{BeKoPr19-4}.

 To evaluate correlators note that the  operators  $u_2 {\q}_1 u^\dagger_2, \q_4 $  contribute only diagonal elements into  (\ref{vectors1},\ref{vectors2}). In the case  of  KIC model this means that only the spin  combinations,   $\Sigma_0=\sigma^y_0 \sigma^j_{1}$,   $\Sigma_n=\sigma^i_n \sigma^z_{n+1}$ might have non-trivial correlations. The corresponding  vectors are given by:
 \begin{equation}
     \Phi_{\sigma_y,\sigma_z}=\bar{\Phi}_{\sigma_y,\sigma_z}=\Phi_0\cos 2h , \label{vectorKIC1}
 \end{equation}
  \begin{equation}
     \Phi_{\sigma_x,\sigma_z}=\bar{\Phi}_{\sigma_y,\sigma_x}=-\Phi_0\sin 2h  ,  \label{vectorKIC2}
 \end{equation}
with $\Phi_0=\frac{1}{2}(1,-1,-1, 1)^T$ being the eigenvector of ${\T}$ for the eigenvalue $\cos 2h $. All other combinations of $x,y,z$ give rise to  zero vectors. After inserting  (\ref{transferKIC},\ref{vectorKIC1},\ref{vectorKIC2}) into  (\ref{four_point_corr}) we  obtain 
\begin{equation}
C_T=\mathcal{C}\cos^T 2h,\label{main_result43}
 \end{equation}
 where prefactors  $\mathcal{C}=
  \mathcal{C}
 (\sigma_0^i\sigma_1^j,\sigma_n^k \sigma_{n+1}^m)$ are given by
 \begin{equation}
\mathcal{C}(\sigma_0^y\sigma_1^z,\sigma_n^y \sigma_{n+1}^z)=1,\qquad  \mathcal{C}(
\sigma_0^y\sigma_1^x,\sigma_n^x \sigma_{n+1}^z
) =\tan^2 2h
 \end{equation}
 \begin{equation}
 \mathcal{C}
 (\sigma_0^y\sigma_1^x,\sigma_n^y \sigma_{n+1}^z)= 
 \mathcal{C}(\sigma_0^y\sigma_1^z,\sigma_n^x \sigma_{n+1}^z) =-\tan 2h
 \end{equation} while   zeroes for all other spin combinations. 
 The same result can be also obtained straightforwardly from the general result  (\ref{Morecorrelators1},\ref{Morecorrelators1}) on  the DFTC model.
 
 The correlator (\ref{main_result43}) decays exponentially for any value of $h$ except for  the set of integrable  points  $h=\frac{1}{4}\pi k, k\in \mathbb{Z}$, where the subleading eigenvalue of $\T$ has  absolute value one.   
In particular, for $h=0$  the   correlator  (\ref{main_result}) vanishes everywhere except for  the cone  border, where it remains constant and does not decay (also for $N<T$). For $h=\pi/4$ the    correlator  vanishes everywhere except for revivals at $N=T$.

\subsection{Two-point correlator in non-dual  KIC model}

While the two point correlator of strictly local operators vanishes everywhere in the dual regime it stays finite as soon as $J-\pi/4=\Delta J \neq 0$.  Below we evaluate      
\begin{equation}C_{x}^x(n=T,  J)=\Trace \left(U_{ J}^{-T} \sigma_n^x {U}_{ J}^{T} \sigma_1^x\right),
\end{equation}
 to the leading order of $\Delta J$:
 \begin{equation}
 C_{x}^x(n=T,  J)
 = C_1 \Delta J + (C_2/2) (\Delta J)^2 +O\left((\Delta J)^4\right).
 \end{equation}
 As $C_1=0$, we need to evaluate
 \begin{equation}
 C_{2}=\frac{d^2}{dJ^2}\Trace \left(U_{ J}^{-T} \sigma_n^x {U}_{ J}^{T} \sigma_1^x\right)\Big |_{J=\pi/4}.
 \end{equation}
 A straightforward calculation gives
 \begin{equation}
     C_2=2\sum_{k=0}^{m-1}\sum_{m=1}^{T-1}\Trace \left([\HI^0(k), \sigma_1^x][\HI^0(m) \sigma_n^x(T)]\right)+
     \sum_{m=1}^{T}\Trace \left([\HI^0(m), \sigma_1^x][\HI^0(m) \sigma_n^x(T)]\right),
 \end{equation}
 where 
 \[\sigma_n^x(T)=U^{-T} \sigma_n^x U^T,\qquad \HI^0(m)=U^{-m}\left(\sum_{n=1}^N \sigma_n^z \sigma_{n+1}^z\right)U^m.\]
 For $n=T$ only the first  term in the above  sum provides the  non-trivial contribution:
 \begin{equation}
     C_2=2\Trace \left([\HI^0, \sigma_1^x][\HI^0 (n-1), \sigma_n^x(n)]\right)
 =8  \Trace \left(U^{-n}\sigma_n^y \sigma_{n+1}^z {U}^{n}\sigma_0^y \sigma_{1}^z \right)=8\cos^n 2h.
 \end{equation}
 
 \end{document}